\documentstyle[aps,prb,epsfig,multicol]{revtex}
\begin{document}
\draft
\title{Shape and Motion of Vortex Cores in Bi$_2$Sr$_2$CaCu$_2$O$_{8+\delta}$}
\author{B.W.~Hoogenboom \cite{email}, M.~Kugler, B.~Revaz \cite{addressRevaz},
I.~Maggio-Aprile, and \O.~Fischer}
\address{DPMC, Universit\'e de Gen\`eve, 24 Quai Ernest-Ansermet,
1211 Gen\`eve 4, Switzerland}
\author{Ch.~Renner}
\address{NEC Research Institute, 4 Independence Way, Princeton,
New Jersey 08540, USA}
\date{\today}
\maketitle
\begin{abstract}
We present a detailed study on the behaviour of vortex cores in
Bi$_2$Sr$_2$CaCu$_2$O$_{8+\delta}$ using scanning tunneling
spectroscopy. The very irregular distribution and shape of the
vortex cores imply a strong pinning of the vortices by defects and
inhomogeneities. The observed vortex cores seem to consist of two
or more randomly distributed smaller elements. Even more striking
is the observation of vortex motion where the vortex cores are
divided between two positions before totally moving from one
position to the other. Both effects can be explained by quantum
tunneling of vortices between different pinning centers.
\end{abstract}

\begin{multicols}{2}
\narrowtext


\section{introduction}
The study of the vortex phases in high temperature superconductors
(HTS's) has lead to both theoretical predictions of several novel
effects and experiments accompanied by challenging
interpretations. The reasons are multiple. First, the
unconventional symmetry of the order parameter --- most likely
$d_{x^2-y^2}$ --- leads to the presence of low-lying quasiparticle
excitations near the gap nodes, which in turn has inspired the
predictions of a nonlinear Meissner effect \cite{Yip:1992}, of a
$\sqrt{H}$ dependence of the density of states at the Fermi level
near the vortex cores ($N(0,\bf{r})$) \cite{Volovik:1993}, and of
a four-fold symmetry of the vortices
\cite{Berlinsky:1995,Salkola:1996,Franz:1998}. However, the
experimental evidence for the first two effects is still
controversial \cite{Moler:1994,Maeda:1995,Revaz:1998,Amin:1998},
and scanning tunneling spectroscopy (STS) measurements on vortex
cores have not shown any clear signature of a $\sqrt{H}$
dependence of $N(0,\bf{r})$ \cite{Maggio:1995,Renner:1998}.
Concerning the four-fold symmetry, a tendency of square vortices
was found in previous measurements \cite{Renner:1998}, but
inhomogeneities make it difficult to be decisive about it.
Interestingly, a four-fold symmetry has been observed around
single atom zinc impurities in Bi$_2$Sr$_2$CaCu$_2$O$_{8+\delta}$
(BSCCO), which are of a smaller size than the vortex cores
\cite{Pan:2000}.

A second reason is related to the interaction of vortices with
pinning centers, responsible for the rich vortex phase diagram of
the HTS's \cite{Blatter:1994}. The pinning of vortices is mainly
due to local fluctuations of the oxygen concentration
\cite{Tinkham:1988,Tinkham:1996,Li:1996,Erb:1999}, and facilitated
by their highly 2D "pancake" character. In BSCCO the areal density
of oxygen vacancies per Cu-O double layer is in fact surprisingly
large: $10^{17}$~m$^{-2}$ \ \cite{Li:1996}, corresponding to an
average distance between the oxygen vacancies of the order of
10~\AA. This vortex pinning results for BSCCO in the absence of
any regular flux line lattice at high fields, as demonstrated by
both neutron diffraction \cite{Cubitt:1993} and STS
\cite{Renner:1998} experiments. Moreover, since the distance
between the oxygen vacancies is of the order of the vortex core
size, one may expect that not only the vortex distribution, but
also the vortex core shape will be dominated by pinning effects,
and {\em not} by intrinsic symmetries like that of the order
parameter. A detailed understanding of the interaction of vortices
with pinning centers will thus be of importance to explain the
lack of correspondence between theoretical predictions about the
vortex shape and STS measurements.

Third, and again as a consequence of the anisotropy of the order
parameter, low-energy quasiparticles are not truly localized in
the vortex core (contrarily to the situation in $s$-wave
superconductors \cite{Caroli:1964,Hess:1989,Renner:1991}). For
pure $d$-wave superconductors these quasiparticles should be able
to escape along the nodes of the superconducting gap. Thus in the
vortex core spectra one expects a broad zero-bias peak of
spatially extended quasiparticle states \cite{Franz:1998}.
However, tunneling spectra of the vortex cores in
YBa$_2$Cu$_3$O$_{7-\delta}$ (YBCO) showed two clearly separated
quasiparticle energy levels, which were interpreted as a signature
of localized states \cite{Maggio:1995}. In BSCCO two weak peaks
have been observed in the some vortex core spectra
\cite{Hoogenboom:2000,Pan:1999}, suggesting a certain similarity
to the behaviour in YBCO. Another important characteristic of
HTS's that follows from the STS studies mentioned above, is the
extremely small size of the vortex cores in these materials. The
large energy separation between the localized quasiparticle states
directly implies that the vortex cores in YBCO are of such a size
that quantum effects dominate. This is even more true in BSCCO:
not only the in-plane dimensions of the vortex cores are smaller
than in YBCO and become of the order of the interatomic distances
\cite{Renner:1998}, but due to the extreme anisotropy of the
material also their out-of-plane size is strongly reduced. This
highly quantized character of vortices in HTS's is equally
demonstrated by the non-vanishing magnetic relaxation rate in the
limit of zero temperature, attributed to quantum tunneling of
vortices through the energy barriers between subsequent pinning
centers \cite{Blatter:1994}.

In this paper we present a detailed STS study of the shape of the
vortices in BSCCO. We will show that this shape is influenced by
inhomogeneities. The samples presented here, which we characterize
as moderately homogeneous, are used to study the behaviour of
vortex cores under these conditions. Apart from the vortex core
shape, this also includes the evolution in time of the vortices.
We will show that both effects can be related to tunneling of
vortices between different pinning centers. This is another
indication of the possible extreme quantum behaviour of vortex
cores in HTS's. A corollary of this paper is that only extremely
homogeneous samples will show intrinsic shapes of vortex cores.

\section{Experimental Details}
The tunneling spectroscopy was carried out using a scanning
tunneling microscope (STM) with an Ir tip mounted perpendicularly
to the (001) surface of a BSCCO single crystal, grown by the
floating zone method. The crystal was oxygen overdoped, with $T_c
= 77$~K, and had a superconducting transition width of 1~K
(determined by an AC susceptibility measurement). We cleaved {\em
in situ}, at a pressure $< 10^{-8}$~mbar, at room temperature,
just before cooling down the STM with the sample. The sharpness of
the STM tip was verified by making topographic images with atomic
resolution. Tunneling current and sample bias voltage were
typically 0.5~nA and 0.5~V, respectively. We performed the
measurements at 4.2~K with a low temperature STM described in
Ref.~\onlinecite{Renner:1990,Kent:1992}, and those at 2.5~K with a
recently constructed $^3$He STM \cite{Kugler:2000}. A magnetic
field of 6~T parallel to the $c$-axis of the crystal was applied
after having cooled down the sample. The measurements presented
here were initiated 3 days after having switched on the field.

The $dI/dV$ spectra measured with the STM correspond to the
quasiparticle local density of states (LDOS). In the
superconducting state one observes two pronounced coherence peaks,
centered around the Fermi level, at energies $\pm\Delta_p$. The
gap size $\Delta_p$ varied from 30-50~meV. In the vortex cores the
spectra are remarkably similar to those of the pseudogap in BSCCO
measured above $T_c$ \ \cite{Renner:1998}, with a total
disappearance of the coherence peak at negative bias, a slight
increase of the zero bias conductivity, and a decrease and shift
to higher energy of the coherence peak at positive bias. To map
the vortex cores we define a gray scale using the quotient of the
conductivity $\sigma(V_p)=dI/dV(V_p)$ at a negative sample voltage
$V_p=-\Delta_p/e$ and the zero bias conductivity
$\sigma(0)=dI/dV(0)$. Thus we obtain spectroscopic images, where
vortex cores appear as dark spots. Since we measure variations of
the LDOS, which occur at a much smaller scale (the coherence
length $\xi$) than the penetration depth $\lambda$, we can get
vortex images at high fields. A tunneling spectrum is taken on the
time scale of seconds, spectroscopic images typically take several
hours (about 12~hours for the images of 100x100~nm$^2$ presented
below). The images therefore necessarily reflect a time-averaged
vortex density.

In all large-scale images we have suppressed short length-scale
noise by averaging each point over a disk of radius $\sim 20$~\AA.
When zooming in to study the shape of individual vortices, we
strictly used raw data. Further experimental details can be found
in previous publications
\cite{Renner:1998,Renner:1998a,Renner:1995}.

\section{Results}

\subsection{Vortex Distribution}
\begin{figure}
 \epsfxsize=70mm
 \centerline{\epsffile{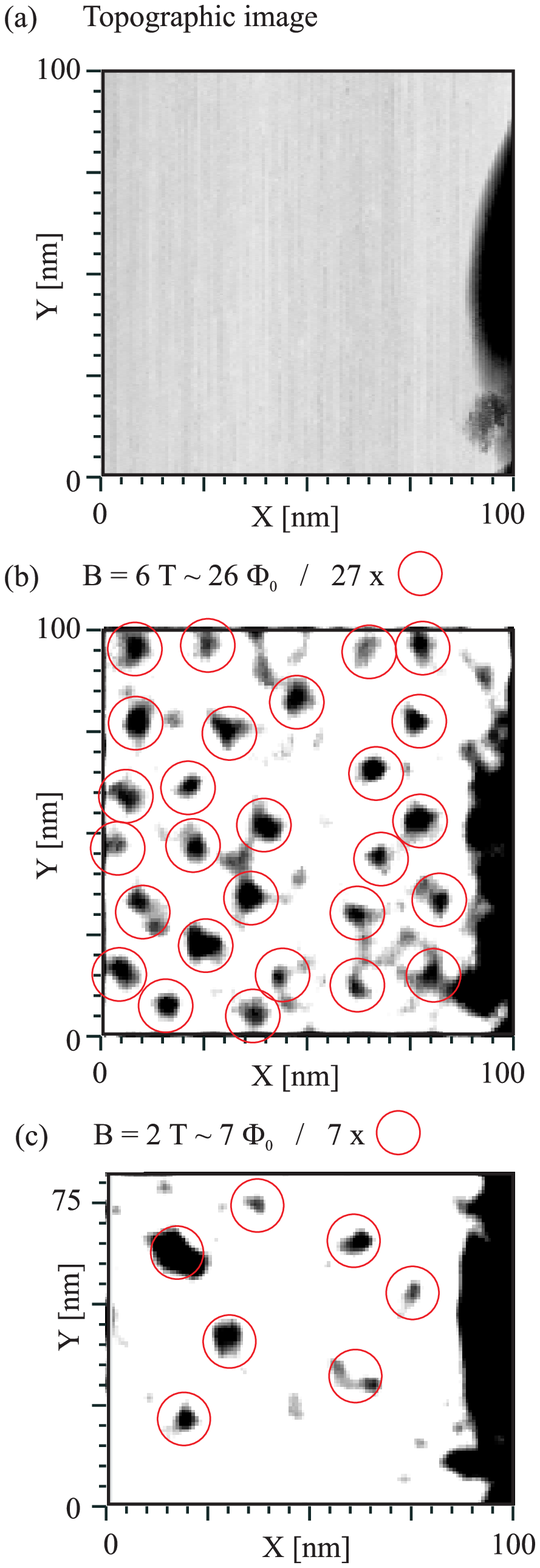}}
 \caption{
 (a)~100x100~nm$^2$ topographic image of the BSCCO surface at 4.2~K and
 6~T, taken at high bias voltage, $V_{bias}=0.4$~V, $I_t=0.6$~nA.
 The structure at the right gradually increases in height from
 0~nm (lightest gray) to almost 5~nm (fully black at the right
 border), and is thus much larger than any atomic details. The
 surface roughness of the gray part is about 1~\AA.
 (b)~Spectroscopic image of the same area, taken simultaneously with~(a);
 dark spots correspond to vortex cores, the dark region at the
 right corresponds to the degraded surface of the structure already observed
 in~(a). For the surface excluding this topographic structure one should
 expect 26 vortices, the image contains 27 (the circles around the vortex
 cores serve as a guide to the eye).
 (c)~Part of the same region, at 2~T, image taken after all measurements at
 6~T. The number of vortices in the image
 again perfectly corresponds to what one should expect for the given surface.}
 \label{greatimage}
\end{figure}

In Fig.~\ref{greatimage} we show spectroscopic images of the
surface of a BSCCO crystal, at different magnetic field strengths.
The large dark structure, clearly visible at the right of
Figs.~\ref{greatimage}(b) and (c), corresponds to a degraded
region resulting from a large topographic structure, already
observed in the topographic image Figs.~\ref{greatimage}(a). The
presence of this structure allows an exact position determination
throughout the whole experimental run. As can be seen in
Fig.~\ref{greatimage}, the number of vortices at 6 and at 2~T, in
exactly the same region, scales very well with the total number of
flux quanta ($\Phi_0$) that one should expect at these field
strengths. This clearly proves that the observed dark spots are
directly related to vortex cores, and not to inhomogeneities,
defects or any form of surface degradation. The large spot in the
upper left corner of Fig.~\ref{greatimage}(c) forms an exception:
it appeared after a sudden noise on the tunnel current while we
were scanning on that position, showed semiconducting spectra
(typical for degraded tunneling conditions) afterwards, and
remained even after having set the external field to 0~T. One
should however not exclude that a vortex is pinned in this
degraded zone. Finally, the size and density of the vortices are
fully consistent with previous measurements
\cite{Renner:1998,Pan:1999}.

Instead of a well ordered vortex lattice, one observes patches of
various sizes and shapes scattered over the surface. This clearly
indicates the disordered nature of the vortex phase in BSCCO at
high fields, again in consistency with previous STM studies
\cite{Renner:1998,Pan:1999} and neutron scattering data
\cite{Cubitt:1993}, and stressing the importance of pinning for
the vortex distribution.

\subsection{Vortex Shapes}
\begin{figure}
 \epsfxsize=70mm
 \centerline{\epsffile{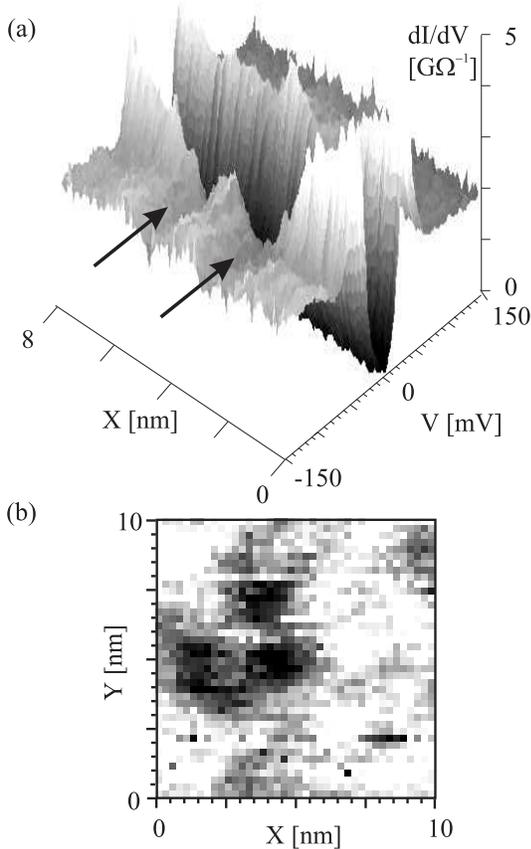}}
 \caption{(a) Spectra along a trace through a vortex core ($B = 6$~T,
   $T = 2.5$~K) reveal that in between regions with vortex core-like
   spectra (indicated by the two arrows) the superconducting coherence
   peaks come up again.
   (b) Image of a vortex core consisting of several separate elements.
   ($B = 6$~T, $T = 4.2$~K).}
 \label{closeup}
\end{figure}

As a next step we increase the spatial resolution in order to
investigate individual vortex cores. Some vortices appear with
square shapes, but most vortices in this study have irregular
shapes. Closer inspection of the tunneling spectra reveals small
zones inside the vortex core that show superconducting behaviour.
That is, when scanning through a vortex core one often observes
(slightly suppressed) coherence peaks (Fig.~\ref{closeup}(a)),
typical for the superconducting state, at some spots {\em inside}
the vortex core. The latter is generally characterized by the {\em
absence} of these peaks. In some cases, the vortex cores are even
truly split into several smaller elements (Fig.~\ref{closeup}(b)),
totally separated by small zones showing the rise of coherence
peaks. This has been verified by measuring the full spectra along
lines through the vortex core, as in Fig.~\ref{closeup}(a).

The smaller elements of a split vortex core  cannot be related to
separate vortices: first, the vortex-vortex repulsion makes it
highly improbable that several vortex cores are so close to each
other; second, counting all these elements as a flux quantum in
Fig.~\ref{greatimage}(b), one finds a total flux through the
surface that is far too large compared to the applied field. One
should note here that the magnetic size of a flux line is of the
order of the penetration depth $\lambda$, two orders of magnitude
larger than the vortex {\em core} splitting observed here.

\subsection{Vortex Motion}
\begin{figure}
 \epsfxsize=90mm
 \centerline{\epsffile{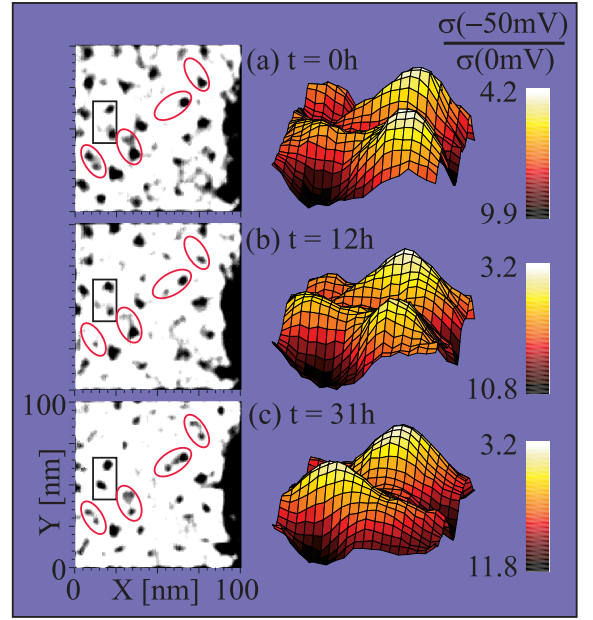}}
 \caption{Sequence of images (each taking about 12~hours) to
 study the behaviour of the vortex cores in time, $B = 6$~T, $T =
 4.2$~K. t corresponds to the starting time of each image.
 (a) t~=~0; (b) t~=~12h; (c) t~=~31h.
 Left: 2D representation. Right: 3D images of the zone marked by the
 rectangles in the 2D images. The vortex core seems to
 be split in (b), before it totally  moves from one position in (a)
 to the other in (c).}
 \label{vortexcreep}
\end{figure}

With subsequent spectroscopic images like
Fig.~\ref{greatimage}(b), one can also study the vortex
distribution as a function of time. We expect the vortex motion to
be practically negligible, since we allowed the vortices to
stabilize for more than 3 days \cite{VanDalen:1996}. However, in
Fig.~\ref{vortexcreep} one can see that many vortices still have
not reached totally stable positions. Many of them roughly stay on
the same positions over the time span of our measurement, but
others move to neighboring positions. Five different cases of
moving vortices are indicated by the ellipses and the rectangle in
Fig.~\ref{vortexcreep}.

In the panels on the left side the precise intensity of each point
is difficult to read out directly. In order to investigate more
quantitatively the time evolution of the vortex distribution, from
one frame to the next, we show in the right part of
Fig.~\ref{vortexcreep} 3D representations of the area that is
marked by the rectangle in the 2D spectroscopic images. They give
an idea of the gray scale used in the 2D images, and provide a
detailed picture of the movement of the vortex core in front, from
the right in Fig.~\ref{vortexcreep}(a) to the left in
Fig.~\ref{vortexcreep}(c). The vortex core at the back does not
move, and serves as a reference for the intensity. We remind that
the intensity, or height in the 3D images, is a measure of the
LDOS, which in a vortex core is different from the superconducting
DOS. It is most interesting to see what happens in
Fig.~\ref{vortexcreep}(b): the (moving) vortex core is {\em
divided} between two positions. Thus, the vortex core moves from
one position to the other, passing through an intermediate state
where the vortex splits up between the two positions. Note that
these two positions do not correspond to two vortices. In fact,
the split vortex is characterized by the lower intensity compared
to the nearby (reference) vortex. This means that the coherence
peak at negative voltage does not completely disappear, as it
should if we had a complete and stable vortex at each of these
positions. Note also that the density of vortices around the
rectangular area on the left side in Fig.~\ref{vortexcreep} will
clearly be too high if we count the mentioned positions and all
positions in the ellipses as individual flux quanta. The split
vortex discussed here is not a unique example. Similar behaviour
can be found for several other vortex cores, as indicated by the
ellipses in Fig.~\ref{vortexcreep}. This gradual change of
position is in striking contrast to the STS observations of moving
vortices in NbSe$_2$ \ \cite{Renner:1993,Troyanovski:1999} and
YBCO \cite{Maggio:1997}.

\subsection{Temperature Dependence}
We performed measurements both at 4.2 and at 2.5~K, on samples cut
from the same batch of crystals. The data taken at 2.5~K (see also
Fig.~\ref{closeup}) are fully consistent with the presented work
at 4.2~K. In Fig.~\ref{2K_data} we provide a general view of the
vortex cores at 2.5~T, including an analogue of the moving vortex
core of Fig.~\ref{vortexcreep}. Though it is hard to obtain any
quantitative data, one can conclude that the vortex cores roughly
have the same size, similar irregular shapes, and examples of
split vortex cores can be easily found.

\section{Discussion}

\subsection{Experimental Considerations}
The observation of such a highly irregular pattern of vortex
cores, as presented above, requires a careful analysis of the
experimental setup. However, the fact that keeping exactly the
same experimental conditions the number of vortex cores scales
with the magnetic field, is a direct proof of the absence of
artificial or noise-related structures in the spectroscopic
images. Furthermore, since topographic images showed atomic
resolution, there is no doubt that the spatial resolution of the
STM is largely sufficient for the analysis of vortex core shapes.

The stability of the magnetic field can be verified by counting
the number of vortices in the subsequent images at 6~T
(Fig.~\ref{vortexcreep}). Since, excluding the split vortices
marked by the ellipses, this number is constant ($26\pm3$), we can
exclude any substantial long time-scale variation of the magnetic
field. Some variation in the total black area from one image to
the other can be related to the tunneling conditions: a little
more noise on the tunnel current will give a relatively large
increase of the small zero-bias conductance. Since we divide by
the zero-bias conductance to obtain the spectroscopic images, this
may lead to some small variations in the integrated black area of
the images.

\subsection{Delocalization}
\begin{figure}
 \epsfxsize=80mm
 \centerline{\epsffile{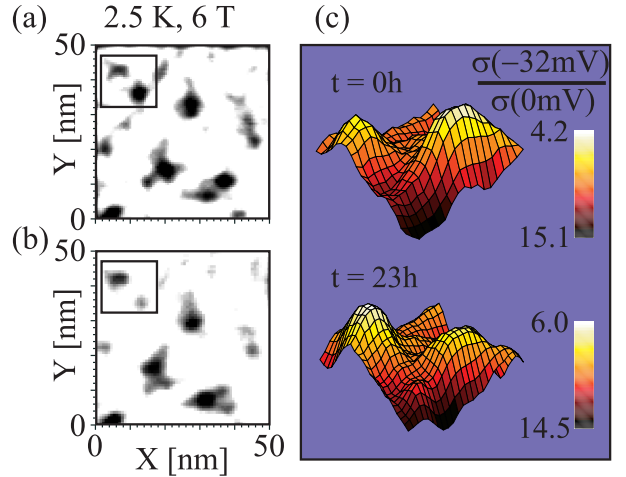}}
 \caption{Subsequent images ((a) and (b)) at $B = 6$~T, $T = 2.5$~K. In (c) a
 3D representation of the square marked in (a) and (b). At 2.5~K one observes
 the same phenomena as at 4.2~K in Fig.~\ref{closeup} and
 Fig.~\ref{vortexcreep}.}
 \label{2K_data}
\end{figure}

Keeping in mind the randomness of the vortex distribution at 6~T
due to pinning of vortices, we now relate both the split vortex
cores (Fig.~\ref{closeup}) and the intermediate state between two
positions (Fig.~\ref{vortexcreep} and Fig.~\ref{2K_data}) to the
same phenomenon: the vortex cores appear to be delocalized between
different positions which correspond to pinning potential wells,
and during the measurement hop back and forth with a frequency
that is too high to be resolved in this experiment. According to
this analysis not only the distribution, but also the observed
shape of the vortex cores is strongly influenced by pinning.

The pinning sites most probably result from inhomogeneities in the
oxygen doping, which are thought to be responsible for the
variations of the gap size (see experimental details). The
distance over which the vortices are split corresponds to the
average spacing between oxygen vacancies ($10-100$~\AA \
\cite{Li:1996}). We did not observe any sign of resonant states
related to impurities, as in recent STM experiments on BSCCO
\cite{Yazdani:1999,Hudson:1999}. The driving forces causing vortex
movements in Fig.~\ref{vortexcreep} and Fig.~\ref{2K_data} are
most probably due to a slow variation of the pinning potential,
resulting from the overall rearrangement of vortices.

The vortex delocalization and movement presented here can directly
be connected to the vortex creep as measured in macroscopic
experiments, like magnetic relaxation \cite{Blatter:1994}. The
main difference, of course, is that we do not observe whole
bundles of vortices moving over relatively large distances, but
only {\em single} vortex cores that are displaced over distances
much smaller than the penetration depth $\lambda$. That is, it
will not be necessary to displace whole groups of vortices, many
of which might be pinned much stronger than the delocalized
vortices we observe. A second difference is the absence of a
uniform direction of the movements in the STM images, most
probably because the Lorentz driving forces have been reduced to
an extremely small value (which also follows from the very gradual
changes in Fig.~\ref{vortexcreep} and Fig.~\ref{2K_data}).

\subsection{Thermal Fluctuations versus Quantum Tunneling}
Regarding now the mechanism responsible for the vortex
delocalization, the main question is whether we are dealing with
thermal fluctuations, or quantum tunneling between pinning
potential wells. In fact magnetic relaxation measurements on BSCCO
show a crossover temperature from thermal to quantum creep of
$2-5$~K \cite{VanDalen:1996,Prost:1993,Aupke:1993,Monier:1998},
which means that with these STM measurements we are on the limit
between the two.

In the case of thermally induced motion, there is a finite
probability for the vortex to jump {\em over} the energy barrier
between the two potential wells. The vortex is continuously moving
from one site to the other, with a frequency that is too high to
be resolved by our measurements. In the case of quantum tunneling,
the vortex is truly delocalized. That is, the vortex can tunnel
{\em through} the barrier, and one observes a combination of two
base states (i.e. positions), like in the quantum text book
example of the ammonia molecule \cite{Feynman:1965}. Thermal
fluctuations will lead to a continuous dissipative motion
\cite{Blatter:1994,Bardeen:1965} between the two sites; quantum
tunneling gives a dissipationless state in which the vortex is
{\em divided} between two positions.

An instantaneous observation of several base states of a quantum
object would be impossible, since each measurement implies a
collapse of the quantum wave function into one state. However, the
STM gives only time averaged images, and with the tunneling
current in this experiment we typically detect one electron per
nanosecond. If the vortex core relaxes back to its delocalized
state on a time scale smaller than nanoseconds, the vortex can
appear delocalized in the STM images. Moreover, it should be clear
that the long time (~12 hours) between the subsequent images in
Fig.~\ref{vortexcreep} and \ref{2K_data} has nothing to do with
the vortex tunneling time; it is tunneling of the vortex that
allows the intermediate state. The creep of vortices (either by
quantum tunneling or by thermal fluctuations) is a slow phenomenon
here. At a given region the pinning potential due to
inhomogenieties and interactions with other vortices evolves on a
time scale of hours, shifting the energetically most favorable
position from one site to the other. However, the tunneling occurs
much faster ($<$ns) than this slow potential evolution, creating
the possibility of a superposition of both states (positions),
each with a probability depending on the local value of the pining
potential. Following the analogy of the ammonia molecule: a moving
vortex core corresponds to an ammonia molecule submitted to an
external external field ("overall pinning potential") that is
slowly changing. Initially, due to this external field the state
with the hydrogen atom at the left is more favorable, then the
field changes such that left and right are equally stable ("split
vortex"), and finally the right state is most probable: the
hydrogen atom has moved from left to right on a long time scale,
whereas the tunneling itself occurs on a much faster time scale.
By the way, since the change of the local pinning potential
results from a rearrangement of all surrounding vortices (over a
distance $\sim\lambda$), the time scale of this change will still
be strongly dependent on the (short) tunneling time itself.

In order to discuss the possibility of quantum tunneling, one
should consider the tunneling time and the for our measurements
negligible temperature dependence
\cite{Blatter:1994,Blatter:1993}. The importance of the tunneling
time is two-fold: the tunneling rate is strongly dependent on the
time needed to pass through the pinning barriers (an effect which
will be extremely difficult to measure directly); and in our
measurements the tunneling time must be faster than the probe
response time (as explained in the previous paragraph). From
collective creep theory \cite{Blatter:1994} one can get an
order-of-magnitude estimate for the tunneling time
$t_c\sim\hbar/U(S_E/\hbar)\sim10^{-11}$~s, with the Euclidian
action for tunneling $S_E/\hbar\sim10^2$ and the effective pinning
energy $U\sim10^2$~K derived from magnetic relaxation measurements
\cite{Blatter:1994,Li:1996,VanDalen:1996,Prost:1993,Aupke:1993,Monier:1998}.
This is clearly below the upper limit of 1~ns set by the probe
response time.

For a discussion about the implications of the temperature
independence of quantum tunneling, one should first consider the
temperature dependence expected for vortex movements that result
from thermal fluctuations. Thermally induced hopping between
different pinning sites should be proportional to $\exp(U/k_BT)$,
where $U$ is again the effective pinning energy
\cite{Blatter:1994}. From magnetic relaxation measurements on
BSCCO one can derive a value of about $10-10^3$~K for this
quantity
\cite{Li:1996,VanDalen:1996,Prost:1993,Aupke:1993,Monier:1998}.
Assuming for the moment that this $U$ determines the hopping of
individual vortices, it should then be compared to the Euclidian
action for quantum tunneling, which with magnetic relaxation
measurements is estimated to be $S_E/\hbar \sim 10^2$ \
\cite{Blatter:1994}, and plays a role like $U/k_BT$ in the
Boltzman distribution. For measurements presented here, it is
important to note again that they were taken more than 3 days
after having increased the field from 0 to 6~T. Since for $B =
6$~T the induced current density $j$ relaxes back to less than
$0.01$ of its initial value in about 10 seconds
\cite{VanDalen:1996}, we are clearly in the limit where $j$ and
thus the Lorentz driving forces (which reduce the energy barrier
for vortex creep) approach zero. This means that the effective
pinning potential $U$ rises, if not to infinity like in isotropic
materials, to a value which in principle is much higher than the
one which determines vortex creep in magnetic relaxation
measurements at comparable field strengths
\cite{Blatter:1994,Tinkham:1996}. With nearly zero Lorentz forces
the tilt of the overall pinning potential will thus be small
compared to the pinning barriers, making thermal hopping {\em
over} the barriers highly improbable (at low temperatures).
Quantum creep, in the limit of vanishing dissipation, is
independent of the collective aspect of $U$, while the probability
for thermal creep decreases as $\exp(-U/k_BT)$ \
\cite{Blatter:1994}. So one can expect quantum creep to become
more important than thermal creep when more time has passed after
having changed the field. In other words, in spite of the fact
that the tilt of the overall pinning potential is small compared
to the pinning barriers, the vortices can still move a little,
i.e. disappearing and reappearing elsewhere.

However, the collective $U$ may be higher than the pinning
barriers for the individual vortex movements observed in our
experiments. Thus, in order to find a lower bound for the latter,
we also estimate $U$ for the moving vortices from our microscopic
measurement. First we calculate the magnetic energy of a vortex
due to the interaction with its nearest neighbors, using
\begin{equation}
  E_{int} = d {{\Phi_0^2} \over {8 \pi^2 \lambda^2}} \sum_i
  \{ln \lgroup {\lambda \over {r_i}} \rgroup +0.12\},
\label{A}
\end{equation}
where $d$ is the length of the vortex segment, $\Phi_0$ is the
flux quantum, $\lambda$ the in-plane penetration depth and $r_i$
the distance to its $i$th neighbor \cite{Tinkham:1996}. Parameters
are conservatively chosen such to give a true minimum estimate for
$E_{int}$ (and thus for $U$): we restrict the out-of-plane extent
of the vortices to zero and thus only take $d = 15$~\AA, the size
of one double Cu-O layer \cite{Harshman:1992} ("pancake
vortices"), and for $\lambda$ take the upper bound of different
measurements, 2500~\AA \
\cite{max_lambda,Martinez:1992,Waldmann:1996}. Taking the vortex
in Fig.~\ref{vortexcreep}(b), and determining the positions
between which it is divided as well as the positions of the
neighboring vortices, one can find the difference between the
magnetic interaction energies of the delocalized vortex at its two
positions. We obtain $E_{int} \sim 120$~K. Now the absence of any
vortex lattice indicates that the pinning potential wells are
generally larger than the magnetic energy difference between the
subsequent vortex positions, and Fig.~\ref{vortexcreep}(b)
reflects a vortex state that is quite common in our measurements
(Fig.~\ref{greatimage}). Following these arguments one can safely
assume that the effective potential well pinning the vortex in
Fig.~\ref{vortexcreep} is larger than this difference: $U >
E_{int} = 120$~K. in agreement with the estimates given above. So
we obtain $U/k_BT > 10-10^2$ for temperatures around 4~K. In the
limit of zero dissipation $S_E/\hbar \sim (k_F\xi)^2$. On the
basis of STS experiments \cite{Maggio:1995,Renner:1998} this can
be estimated to be $\leq 10$. This value is smaller than the one
quoted above, and suggests that quantum tunneling is dominant in
our measurements.

The most direct evidence for quantum creep can be obtained from
measurements at different temperatures. The hopping rate for
thermally induced movements is given by $\omega_0\exp(-U/k_BT)$,
where $U$ is the pinning potential, and $\omega_0$ the
characteristic frequency of thermal vortex vibration
\cite{Tinkham:1996}. Assuming $U = 100$~K, and a conservatively
large estimate of $\omega_0 \sim 10^{11}$~s$^{-1}$, the hopping
rate should drop from 1~s$^{-1}$ to 10$^{-7}$~s$^{-1}$ on cooling
from 4.2 to 2.5~K. This gives a huge difference between the
respective measurements at these temperatures. However,
spectroscopic images at 4.2 and 2.5~K show the same pattern of
moving and delocalized vortices. Following the same kind of
estimations as above, the delocalized vortex at 2.5~K
(Fig.~\ref{2K_data}) gave $U > 210$~K, which makes thermal creep
even more unlikely here. Even if the frequency of the individual
thermal vortex movements were too high to be resolved by our
measurements {\em both} at 4.2 and at 2.5~K (this would mean a
rather unrealistic characteristic frequency $\omega_0 > 10^{15}$),
one would still expect to see a difference. As a matter of fact,
the driving force for the vortex movements results from an overall
rearrangement of vortices. This means that the displacements of
vortices will always depend on the hopping frequency, and that
even for very high hopping rates one should observe a reduction of
the number of vortices that are displaced in our images, when the
hopping rate is reduced by a factor 10$^7$.

\section{Conclusion}
We observed vortex cores that were delocalized over several pinning
potential wells. Regardless of the exact mechanism (thermal
hopping or quantum tunneling) responsible for this delocalization,
our measurements point out that pinning effects not only dominate
the distribution of the vortex cores, but also their shape. As a
consequence intrinsic (four-fold?) symmetries of the vortex cores
will be obscured in microscopic measurements. The delocalization
of the vortex cores implies that the vortex cores in this
study appear larger than their actual --- unperturbed --- size,
indicating a coherence length that is even smaller than was
expected on the base of previous studies \cite{Renner:1998}.

The analysis given above strongly favors an interpretation in
terms of quantum tunneling of vortex cores. This would not only
mean the first microscopic signature of the vortex quantum
tunneling as derived from magnetic relaxation measurements, it is
also a further indication \cite{Arndt:1999} that objects of larger
size and complexity than one or several atoms can appear as a
superposition of different quantum states.

\acknowledgements This work was supported by the Swiss National
Science Foundation.


\end{multicols}
\end{document}